\documentclass[aps,nofootinbib,superscriptaddress]{revtex4}
\usepackage{color,epsfig,graphicx,graphics,amssymb,amsmath,subeqnarray,units,color,subfigure,cancel}

\def\baselinestretch{1.2}

\begin{document}

\title{A two-dimensional model of low-Reynolds number swimming beneath a free surface}
\author{Darren Crowdy}
\affiliation{Department of Mathematics, Imperial College, London, SW7 2AZ, United Kingdom}
\author{Sungyon Lee}
\affiliation{Hatsopoulos Microfluids Laboratory, Department of Mechanical Engineering, Massachusetts Institute of Technology, 77 Massachusetts Avenue, Cambridge, MA 02139, USA}
\author{Ophir Samson}
\affiliation{Department of Mathematics, Imperial College, London, SW7 2AZ, United Kingdom}
\author{Eric Lauga}
\affiliation{Department of Mechanical and Aerospace Engineering, University of California San Diego, 9500 Gilman Drive, La Jolla CA 92093-0411, USA.}
\author{A. E. Hosoi}
\affiliation{Hatsopoulos Microfluids Laboratory, Department of Mechanical Engineering, Massachusetts Institute of Technology, 77 Massachusetts Avenue, Cambridge, MA 02139, USA}

\newcommand\dynpercm{\nobreak\mbox{$\;$dyn\,cm$^{-1}$}}
\newcommand\cmpermin{\nobreak\mbox{$\;$cm\,min$^{-1}$}}

\providecommand\bnabla{\boldsymbol{\nabla}}
\providecommand\bcdot{\boldsymbol{\cdot}}
\newcommand\biS{\boldsymbol{S}}
\newcommand\etb{\boldsymbol{\eta}}

\newcommand\Real{\mbox{Re}} 
\newcommand\Imag{\mbox{Im}} 
\newcommand\Rey{\mbox{\textit{Re}}}  
\newcommand\Ca{\mbox{\textit{Ca}}}  
\newcommand\Pran{\mbox{\textit{Pr}}} 
\newcommand\Pen{\mbox{\textit{Pe}}}  
\newcommand\Ai{\mbox{Ai}}            
\newcommand\Bi{\mbox{Bi}}            

%
%

\newcommand\ssC{\mathsf{C}}    
\newcommand\sfsP{\mathsfi{P}}  
\newcommand\slsQ{\mathsfbi{Q}} 

\newcommand\hatp{\skew3\hat{p}}      
\newcommand\hatR{\skew3\hat{R}}      
\newcommand\hatRR{\skew3\hat{\hatR}} 
\newcommand\doubletildesigma{\skew2\tilde{\skew2\tilde{\Sigma}}}

\newsavebox{\astrutbox}
\sbox{\astrutbox}{\rule[-5pt]{0pt}{20pt}}
\newcommand{\astrut}{\usebox{\astrutbox}}

\newcommand\GaPQ{\ensuremath{G_a(P,Q)}}
\newcommand\GsPQ{\ensuremath{G_s(P,Q)}}
\newcommand\p{\ensuremath{\partial}}
\newcommand\tti{\ensuremath{\rightarrow\infty}}
\newcommand\kgd{\ensuremath{k\gamma d}}
\newcommand\shalf{\ensuremath{{\scriptstyle\frac{1}{2}}}}
\newcommand\sh{\ensuremath{^{\shalf}}}
\newcommand\smh{\ensuremath{^{-\shalf}}}
\newcommand\squart{\ensuremath{{\textstyle\frac{1}{4}}}}
\newcommand\thalf{\ensuremath{{\textstyle\frac{1}{2}}}}
\newcommand\Gat{\ensuremath{\widetilde{G_a}}}
\newcommand\ttz{\ensuremath{\rightarrow 0}}
\newcommand\ndq{\ensuremath{\frac{\mbox{$\partial$}}{\mbox{$\partial$} n_q}}}
\newcommand\sumjm{\ensuremath{\sum_{j=1}^{M}}}
\newcommand\pvi{\ensuremath{\int_0^{\infty}%
 \mskip \ifCUPmtlplainloaded -30mu\else -33mu\fi -\quad}}

\newcommand\etal{\mbox{\textit{et al.}}}
\newcommand\etc{etc.\ }
\newcommand\eg{e.g.\ }

\newcommand{\others}[1]{\textcolor{red}{#1}}

\newcommand{\sungyon}[1]{\textcolor{red}{#1}}
\newcommand{\B}{\begin{equation}}
\newcommand{\E}{\end{equation}}
\newcommand{\Bb}{\begin{equation*}}
\newcommand{\Ee}{\end{equation*}}
\newcommand{\BE}{\begin{eqnarray}}
\newcommand{\ESS}{\end{split}}
\newcommand{\mc}{\mathcal}
\newcommand{\EE}{\end{eqnarray}}
\newcommand{\NN}{\nonumber}
\newcommand{\NNN}{\NN\\\NN\\}
\renewcommand {\baselinestretch}{1.}

\def\di{\displaystyle}

\date{\today}

\begin{abstract}

Biological organisms swimming at low Reynolds number are often influenced by the presence of rigid boundaries and soft interfaces. In this paper we present an analysis of  locomotion  near a free surface with surface tension. Using a simplified two-dimensional singularity model,  and combining a complex variable approach with conformal mapping techniques, we demonstrate that the deformation  of a free surface can be harnessed to produce steady locomotion parallel to the interface. The crucial physical ingredient lies in the nonlinear hydrodynamic coupling between the disturbance flow created by the swimmer and the free boundary problem at the fluid surface.

\end{abstract}

\maketitle

\section{Introduction}

Low-Reynolds number swimming near solid boundaries or interfaces can exhibit interesting and unexpected features.  
In particular, the presence of long-range interactions typical of flows at low Reynolds numbers implies
that, in general, boundary effects can not be ignored \citep{Brennen1977,laugapowers09}.
For instance, \textit{E. coli} cells are observed to change their swimming trajectories from straight to circular when they are moving parallel to a solid surface \citep{Maeda1976,Berg1990,Frymier1997,Lauga2006a}, a  behaviour modification which may have important implications in the formation of biofilms \citep{costerton95}.  The motion of microorganisms near soft interfaces, such as spermatozoa motility through the mucus-filled female reproductive track \citep{Suarez2006}, is even more intriguing 
as the nonlinear coupling between the motion of the swimmer and the changing shape of the interfaces  adds an extra level of complexity to the problem.

The study of low Reynolds number swimmers near a no-slip wall 
has received considerable attention in the past, and we refer to the reviews by \cite{Brennen1977} and \cite{laugapowers09} for a discussion of the relevant literature.  
Most theoretical work has focused on quantifying the change in swimming speed and energetics near solid boundaries \citep{Reynolds1965,katz74,katz75,katzblake,fauci95}.
More recent work has addressed  the dynamics of confined swimmers, and tackled the subtle interplay between the time evolution of a swimmer's orientation and its position.
For example, a well-known feature of swimming   near a solid boundary is that
organisms moving  at low Reynolds number tend to be  attracted to solid surfaces \citep{Rothschild,winet84a,fauci95,cosson03,woolley03, Hernandez-Ortiz2005,Berke}.
This phenomenon can be rationalized by a fundamentally hydrodynamical mechanism
in which the interaction with the rigid boundary causes a swimmer to 
reorient itself in such a way that it is eventually attracted to its hydrodynamic image system in the wall  \citep{Berke}.

Other studies have revealed
additional dynamical features of a swimmer's behaviour near a wall.
\cite{OrMurray} have conducted numerical experiments to understand 
the wall-bounded  dynamics of  model swimmers from a control and dynamical systems
perspective. In addition to the existence of a steady state in which the swimmers travel
in a steady rectilinear motion parallel to the wall,  the authors found that the generic
motion of a swimmer can be described by nonlinear periodic orbits along the wall  with  complicated spatio-temporal structure. These observations have since been corroborated by laboratory
experiments involving small robotic swimmers in a tank of viscous fluid \citep{Zhang}.
Motivated by these studies, \cite{CrowdyOr} have recently proposed a simple two dimensional model of
a swimmer near a wall. Using  a complex variable formulation of the problem, they obtained results in 
agreement with those of \cite{OrMurray} and  
 \cite{Zhang}.

In the current paper, we address the coupling between a low-Reynolds swimmer and a
surface which can deform, and focus on the case of a  free  interface with surface tension.
Previous work considered how the {\em unsteady} deformation of soft surfaces generated by time-reversible flows could  provide new modes of locomotion and pumping  \citep{Trouilloud2008}.  Here we ask the following question: Can a low-Reynolds number body exploit the deformation of a free surface to swim steadily?
To emphasize the role played by the surface deformation,  we consider  swimmers which cannot swim in the absence of a free surface, and determine whether they are able, through the disturbance flow field they are creating and the subsequent surface deformation,  to  acquire locomotive abilities. 
Specifically, using the  modeling approach of \cite{CrowdyOr}, we focus  on identifying a steady mode of locomotion in which a swimmer translates at a constant speed parallel to the undisturbed free surface.
Since, in the neighborhood of a flat no-slip wall, the motion of a swimmer  generically follows a time-dependent periodic orbit,  it is not clear {a priori}  that a steadily translating swimmer  motion beneath a free capillary surface is possible. 
What we show below is that such a mode of locomotion is  indeed possible and, within our model can be described in a mathematically explicit way.

Our study was originally motivated by the discovery of a peculiar mode 
of locomotion employed by water snails that crawl underneath the free surface.  
Separated from the interface by a thin layer of mucus, these organisms 
deform their foot to create a lubrication flow inside the mucus layer.  This flow 
results in  deformations of the free surface, which in turn rectifies the flow, 
allowing the water snails to move \citep{Lee2008}.  The  analysis  in this paper
contained two significant constraints, namely the gap between the 
swimmer and the air-water interface was assumed to be thin, and the 
deformation of the free surface was assumed to asymptotically small.  
The work in the current paper removes these constraints 
and considers a more general mechanism for a swimmer translating steadily beneath the free surface. 

Low Reynolds number swimmers
exert no net force and no net torque  on the flow, and it is  precisely these constraints that dictate the subsequent speed of the
swimmer and its angular velocity. Here we introduce 
a mathematical representation of the swimmer as a two-dimensional torque-free point 
stresslet which, by definition, is force and torque-free.
This type of singularity model 
has been widely used in modeling suspensions of force-free particles \citep{Batchelor1970} 
and swimming microorganisms \citep{Pedley1992,Hatwalne2004,Hernandez-Ortiz2005}. 
The approach is equivalent to considering the swimmer 
on distances much larger than its intrinsic size, so that its precise geometric structure and the fine details of
its swimming protocol are encapsulated in the effective
far-field multipole structure.
The two-dimensional assumption, although idealized and not directly relevant to biological swimmers, allows us to explicitly solve for the nonlinear free boundary problem, thereby shedding light on this new mode of locomotion.

Our mathematical approach is inspired by 
the work of \cite{Jeong1992} who
considered surface deformations generated by
 two counter-rotating cylinders beneath a free surface at low Reynolds numbers.
In a similarly idealized model, the flow generated by the two counter-rotating cylinders is
modeled by
a single potential dipole
located on the axis of symmetry of the deformed free surface.
This flow results in symmetric deformations of the interface which are
calculated, for a given dipole strength and fluid properties, by means of 
conformal mapping methods.  Notably, the conformal map 
approach allowed them to produce {exact} 
solutions even for large nonlinear deformations of the free surface.  
Furthermore, their analytic results exhibit remarkable agreement 
with the experimental data of free surface deformations for different rotation rates of the cylinders. 
In the spirit of \cite{Jeong1992}
we therefore use a conformal map to solve for the shapes of the free surface governed
by the interaction between surface tension and the flow field generated by the swimmer, represented as a combination of singularities.
The difference here is that we must allow for non symmetric deformations of the interface
and adapt the analysis to admit a stresslet singularity in the fluid (rather than a potential dipole).

This paper is organized as follows.  In~\S\,\ref{sec:motiv}, the two
dimensional Stokes equations and relevant boundary conditions
 are introduced in complex variables. The singularity model 
 approach is then explained in~\S\,\ref{sec:sing}.  Section~\ref{MI} uses a method of images
 to demonstrate that steady motion of a swimmer beneath a flat undeformed interface
 is not possible.
Section \,\ref{sec:method} then introduces
a conformal mapping approach that enables us to explore solutions
in which the free surface admits essentially arbitrary deformations. 
Section \,\ref{eff} 
gives a characterization of the class of steadily translating solutions,
 and is followed by the conclusions in~\S\,\ref{sec:concl}.

\section{Complex variable formulation of Stokes flow}\label{sec:motiv}

Let the two-dimensional quiescent fluid occupy the area beneath a deformable fluid-air interface, $D$.  
The fluid is assumed to be incompressible and, in the Stokes r\'egime,
the streamfunction $\hat \psi$ is known to satisfy the biharmonic equation
\B  \nabla^4 \hat \psi(\hat x,\hat y)= 0. \label{eq:biharm} \E \vskip 0.1truein \noindent
Introducing the complex-valued coordinate
$\hat z=\hat x+{\rm i}\hat y$, 
it is possible to write the general solution of the biharmonic equation in the form
\B \hat \psi = \mbox{Im}[ \overline{\hat z}\hat f(\hat z)+\hat g(\hat z)].  \E \vskip 0.1truein \noindent
Here $\hat f\equiv \hat f(\hat z)$ and $\hat g \equiv \hat g(\hat z)$ 
are two functions which must be analytic functions of $\hat z$
inside the fluid region except at isolated points where
singularities are deliberately introduced in order to model particular
flow conditions. These functions are sometimes referred to as
{\em Goursat functions}. 

It is possible \citep{Langlois} to express all the usual physical variables in terms of these two functions.
Indeed, it can be shown that
\B 
\begin{split}
{\hat p \over \mu} - {\rm i} \hat\omega &= 4 \hat f'(\hat z), \\
\hat u + {\rm i} \hat v &= -\hat f(\hat z) + \hat z \overline{\hat f'(\hat z)} + \overline{\hat g'(\hat z)}, \\
\hat e_{11} + {\rm i} \hat e_{12} &= \hat z \overline{\hat f''(\hat z)} + \overline{\hat g''(\hat z)}.
\end{split} \label{eq:eqPHYS}
\E
Here, $\hat p$ is the fluid pressure,
$\hat \omega$ is the vorticity, $(\hat u, \hat v)$ is the fluid velocity
and $\hat e_{ij}$ is the fluid rate-of-strain tensor. The dynamic fluid viscosity is $\mu$.
Primes denote differentiation with respect to $\hat z$, and overbars denote complex conjugates.

The stress boundary condition on the free surface requires that the normal fluid stress is balanced by the surface
tension and that the tangential stress vanishes. This can be written as
\begin{equation}\label{eq:stress_cond}
-\hat p n_i + 2 \mu \hat e_{ij} n_j = \sigma \kappa n_i
\end{equation}
where $\sigma$ is the surface tension, $\kappa$ is the surface curvature, and $n_i$ is the outward unit vector normal to the interface.
In addition, the kinematic condition on the interface requires that the normal velocity of the interface
equals the normal fluid velocity.

The governing equations, Goursat functions, and corresponding boundary conditions are non-dimensionalized as follows:
\B
\begin{split}
\hat z = \hat h z,~~~
&\hat z_{d} =  \hat h z_{d},~~~
\hat u +{\rm i}\hat v = \hat U (u+{\rm i}v),~~~
\hat \psi =  \hat U \hat h \psi,~~~\\
&\hat f =  \hat U f,~~~
\hat g =  \hat U \hat h g,~~~
\hat p = \frac{\mu \hat U}{\hat h} p,
\end{split}
\E
where $\hat z_{d}$ is the dimensional location of the swimmer, $\hat h$ is the magnitude of $\hat z_{d}$ or the vertical distance of the swimmer from the interface, and
$\hat U$ is a characteristic speed of translation.
The capillary number Ca, which reflects the dimensionless ratio of viscous to capillary effects, is defined
as 
\begin{equation}
\mbox{Ca} = {\mu \hat U \over \sigma}.
\end{equation}

\noindent  It can be shown that the complex form of the
stress condition (\ref{eq:stress_cond})  on the air-fluid interface is equivalent to the relation
\B \frac{dH}{ds} = -\frac{{\rm i}}{2\mbox{Ca}}\frac{d^2z}{ds^2},\qquad \qquad\E \vskip 0.1truein \noindent
where $ds$ is a differential element of arc length along the free surface
and 
\B
H \equiv f(z) + z\overline{f}'(\overline{z}) + \overline{g}'(\overline{z}).\E
Hence, the stress condition can be integrated once with respect to $s$ to give
\B f(z) + z\overline{f}'(\overline{z}) + \overline{g}'(\overline{z}) = -\frac{{\rm i}}{2\mbox{Ca}}\frac{dz}{ds},  \qquad \qquad \,\,\,  \label{eq:stressbalance}\E \vskip 0.1truein \noindent
where, without loss of generality,
the constant of integration has been set equal to zero. 

\section{A singularity model of the swimmer}\label{sec:sing}

\begin{figure}
\begin{center}
\includegraphics[scale=0.32]{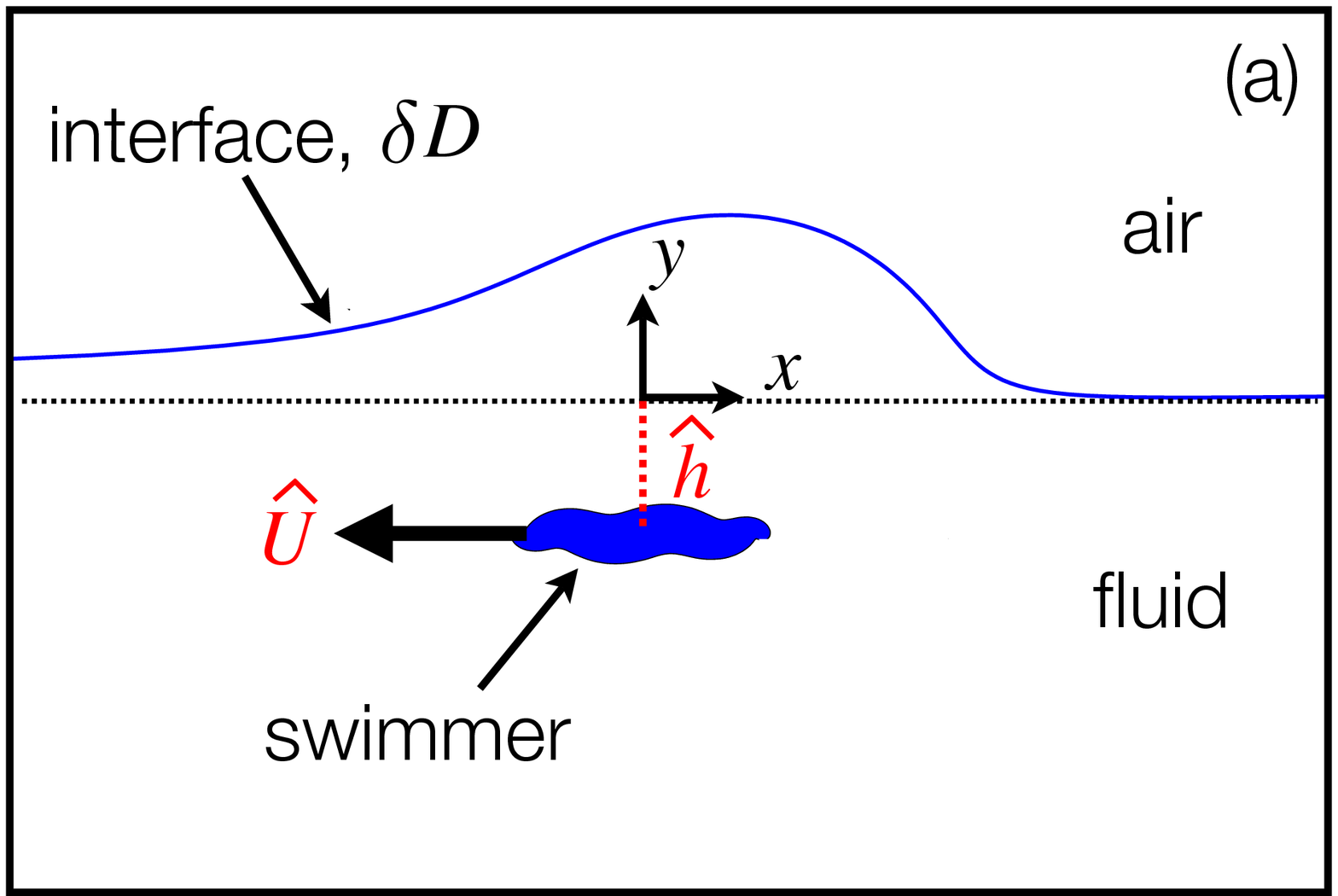}
\hskip .2truein\includegraphics[scale=0.315]{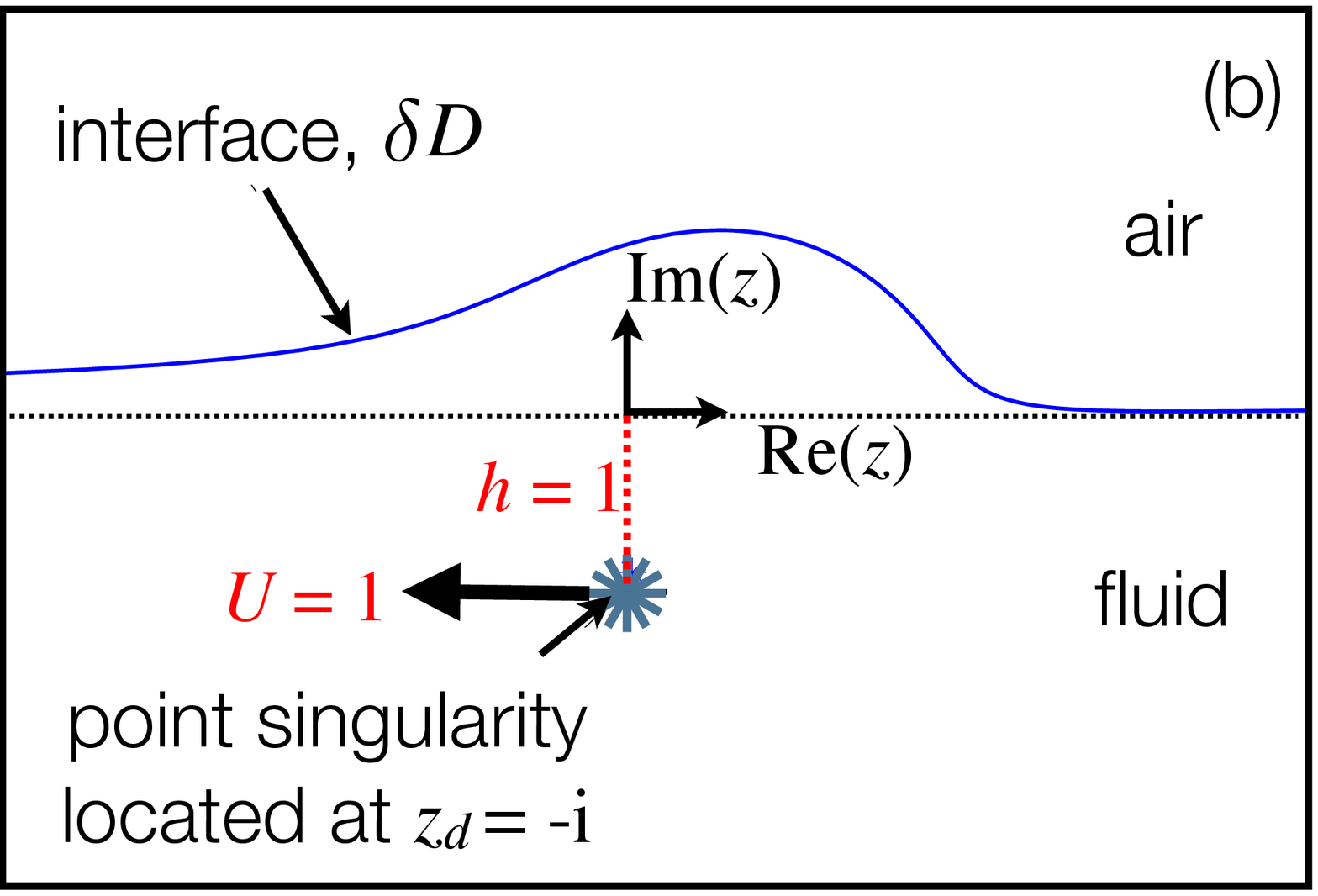}
\caption{Illustration of the   singularity model: A finite-body swimmer beneath a free surface is modelled as a point stresslet singularity with superposed potential dipole and quadrupole. \label{Fig0}}
\end{center}
\end{figure}

From the first equation in ({\ref{eq:eqPHYS}) it is clear that
singularities of the Goursat function $f(z)$ will be related to
local singularities in the pressure field and hence, to
localized force singularities. A {\em logarithmic} singularity of $f(z)$ corresponds to what is
often referred to as a {\em stokeslet},
or point force singularity \citep{Pozrikidis1992} which, as previously discussed, is not allowed
owing to the condition that the swimmer
exerts no net force on the flow. 

The next order singularity (the derivative of the logarithm)
is a simple pole. 
If near $z_d$, $f(z)$ has a simple pole singularity, 
\begin{equation}
 f(z) = {s^* \over z-z_d} + {\rm analytic}, 
 \label{eq:eqFF} 
 \end{equation}
then, in order to ensure that the velocity field scales like $1/|z-z_d|$
 (rather than $1/|z-z_d|^2$) we must also have
\begin{equation}
g'(z) = {s^* \overline{z_d} \over (z-z_d)^2} + {\rm analytic}. 
\label{eq:eqGGG} 
\end{equation}
Thus, if $f(z)$ and $g'(z)$ locally have the behavior 
(\ref{eq:eqFF}) and (\ref{eq:eqGGG}) respectively
near $z_d$ then  there is a {\em stresslet} of strength 
 $s^*$
at $z_d$.
In general, if $g(z)$ has a simple pole near some point $z_d$ of the form
\B g(z) = \frac{d}{z-z_d} + {\rm analytic} \label{eq:eqdip}, \E
then we say there 
is a {\em dipole} of strength $d$ at $z_d$. 
A stresslet singularity of strength $s^*$ at $z_d$ therefore corresponds to a simple pole of $f(z)$ at $z_d$ 
with residue $s^*$ together with a simple pole of $g(z)$ with residue
$-s^* \overline{z_d}$ at the same point.

There is a physical way to understand this singularity of $f(z)$.
Swimmers at low Reynolds numbers propel themselves
by exerting a local force on the flow (for example, the waving flagellum of a spermatozoa) which
is then counterbalanced by a net drag on its body (the head of
the spermatozoa plus its flagellum) leaving the total net force on it
equal to zero \citep[e.g.~see review in][]{laugapowers09}. If this scenario is modeled as two logarithmic singularities of $f(z)$ (two point forces)
drawing infinitesimally close together with equal and opposite strengths tending
to infinity at a rate inversely proportional to their separation, the limit is
precisely a simple pole singularity of $f(z)$ of the form (\ref{eq:eqFF}).

In a general singularity description of a swimmer, 
$g'(z)$ is also singular at $z_d$.
We already know it must have a second order pole (\ref{eq:eqGGG}) (this is
associated with the stresslet) but it can have additional singularities.
The singularities of $g(z)$ are {\em potential} multipoles because, as is clear
from (\ref{eq:eqPHYS}), only the singularities of $f(z)$ contribute to the
vorticity of the flow.
Different swimmers generate different effective singularities
according to their particular swimming protocol. 
Any choice of singularities that we assume $g'(z)$ to have at $z_d$
is therefore a manifestation of our choice of swimmer type.
It is not clear, {\em a priori}, how to pick either the type of these singularities
of $g'(z)$ or their magnitudes.

In this paper we adopt the same singularity model of a swimmer used
by \cite{CrowdyOr} in their studies of a low-Reynolds number swimmer
near a no-slip wall. 
They motivated their choice of singularities by considering a concrete
model of a finite-area
circular ``treadmilling'' swimmer of radius $\epsilon$.
It was supposed that, on its surface, the swimmer generates
a purely tangential surface velocity given by
\begin{equation}
U(\phi,t) = 2 V \sin(2(\phi-\theta(t)))
\end{equation}
where $V$ is a constant (setting the time-scale of the
swimmer's motion), $\phi$ is the angular variable
and $\theta(t)$ is a distinguished angle taken to be the
direction in which the head
of the swimmer is pointed.
By solving a boundary value problem for
the flow associated with this swimmer in an unbounded Stokes flow,  it is possible to show that 
such a swimmer has an effective
singularity description consisting of a stresslet of strength $\mu(t) = \epsilon V \exp(2 {\rm i} \theta(t))$
with a superposed potential quadrupole of strength $2 \mu(t) \epsilon^2$.

This model is a particular case of a general class of
simplified swimmers first considered in a theoretical study due to \cite{Blake1972}
who looked at the effect of imposing velocity profiles of general form on the surface of a circular swimmer. Such an ``envelope model''  captures the
macroscopic effect of the motion of many small-scale beating cilia on the swimmer surface. Similarly, cilia-aided crawling of organisms beneath a free surface has been observed in nature, in particular for some families of snails. 
\cite{Copeland1919} concluded
that the locomotion of {\em Alectrion trivittata}, which
crawls upside down on the surface, relies solely on the ciliary
action. He conducted a similar study on {\em Polinices duplicata}
and {\em Polinices heros}, both of which were observed to use
both cilia and muscle contraction for locomotion on hard
surfaces \citep{Copeland1922}.
Only ciliary motion was employed by the young
{\em Polinices heros} when crawling inverted beneath the surface.

Prompted by the success of the previously described singularity model, we extend the
study to point swimmers (beneath a free surface) within the same general class: that is, a point
stresslet superposed with a  potential dipole and quadrupole. 
The dipole has been included because it is a lower order singularity than the quadrupole
and there is no reason {\em a priori} to suppose it is absent (moreover steady rectilinear
motion is expected to involve a dipole in its singularity description).
This means
that we will seek $f(z)$ and $g(z)$ with the functional forms
\begin{equation}
 f(z) = \frac{s^*}{z+{\rm i}} + f_0 + f_{1}(z+{\rm i}) + \dots, ~~~
g(z) =  {q^* \over(z+{\rm i})^2} + {(- {\rm i} s^* + d^*) \over (z+{\rm i})}  +
 g_0 + g_{1}(z+{\rm i}) + \dots
 \label{eq:eqGForm}
 \end{equation}
 where the singularity is at $z=-{\rm i}$ and $f_0, f_1, g_0$ and $g_1$ are constants.  We will refer to 
 $s^*$ as the stresslet strength, $q^*$ as the quadrupole strength and $d^*$ as the
 dipole strength (note that part of the coefficient of $1/(z+{\rm i})$ in $g(z)$ is naturally associated
 with the stresslet singularity as seen in (\ref{eq:eqFF}) and (\ref{eq:eqGGG})).
 We will not make any {\em a priori} assumptions on the relative magnitudes of
 $s^*, d^*$ and $q^*$ since, for a steady solution, we expect these to be determined
 by the conditions for equilibrium.

\cite{CrowdyOr} show that the evolution equations for the swimmer position $z_d(t)$ and its orientation
$\theta(t)$ are given by the dynamical system
\begin{equation}
{d z_d(t) \over dt} = -f_0 + z_d \overline{f_1} + \overline{g_1}, \qquad
{d \theta(t) \over dt} = - 2 {\rm Im}[f_1].
\end{equation}
The first equation states that the swimmer moves with the finite part of the fluid velocity at the
swimmer position, while the second equation states that its angular velocity equals half the regular part of the vorticity
at the swimmer position.
In the present paper we adopt these same evolution equations but focus on finding
equilibrium solutions in which the swimmer translates steadily in the direction of the undeformed
interface (i.e., parallel to the $x$-axis). In a co-travelling frame we therefore need to find solutions
satisfying the conditions
\begin{equation}
0 = -f_0 + z_d \overline{f_1} + \overline{g_1}, \qquad
0 = {\rm Im}[f_1].
\label{eq:eqEVOL}
\end{equation}
The first equation ensures that the swimmer is stationary in the co-moving frame. The second equation ensures that the local vorticity at the swimmer position vanishes so that its orientation
remains fixed in time. Note that for a no-slip wall the time evolution of the swimmer's orientation
proves to be a crucial ingredient in understanding the dynamics of swimmers near those
surfaces \citep{Berke}, and  the same is expected to be true near a free surface.

\section{Steady swimming beneath a flat interface: method of images \label{MI}}

\begin{figure}
\begin{center}
\includegraphics[scale=0.3]{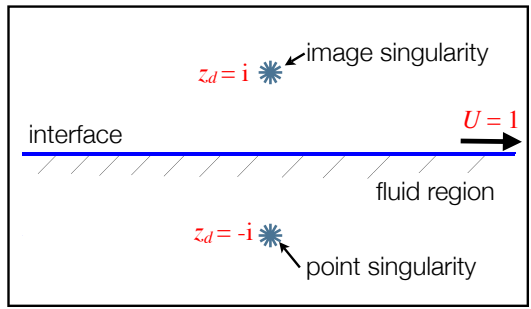}
\caption{The method of images near a flat free surface: to satisfy the boundary condition on the surface an image singularity must be introduced at the reflection of
the point swimmer position.\label{MIImage}}
\end{center}
\end{figure}

It is instructive to first examine whether it is possible to find
a solution for a point swimmer (of the type just described)
translating steadily beneath a flat undeformed interface. To do so
we must seek $f(z)$ and $g(z)$ satisfying the boundary conditions
on the free surface, with singularities
of the form (\ref{eq:eqGForm}), and
satisfying 
conditions (\ref{eq:eqEVOL}).
Owing to the simplicity of the geometry we search for such a solution
using the method of images, as originally introduced by Blake and co-workers 
for three-dimensional flows near no-slip walls \citep{blake71_image,blake74_image}.  
We now demonstrate that such a solution does not exist.
In subsequent sections we succeed in identifying steadily translating
solutions in which the interface deforms.
This highlights the crucial role played by the deformability of the interface
in providing a mechanism for steady translation of the swimmer.

Consider a swimmer moving at a prescribed horizontal speed 
$U=-1$ beneath a flat interface in a Stokes flow 
with no background flow.  It is
convenient to work in a frame traveling with the swimmer.
Then the swimmer is stationary at $z_d=-{\rm i}$ while the fluid in the far-field
has unit speed in the $x$-direction. 

We seek a solution of the form
\B f(z) = \frac{s^*}{z+{\rm i}} + \frac{\hat{s^*}}{z - {\rm i}} + f_0 \label{eq:eqFFF} \E \vskip 0.1truein \noindent
where, in addition to the stresslet of strength $s^*$ at $z_i=-{\rm i}$, we have placed  
an image stresslet of strength $\hat{s^*}$ (to be determined) at $z={\rm i}$ 
and $f_0$ is a constant. Figure \ref{MIImage} shows a schematic of both the swimmer and its image. 
The form of $g'(z)$ is now forced by the stress boundary condition on this interface.
On $\overline{z}=z$ we have, from (\ref{eq:stressbalance}),
\B  \overline{g}'({z}) = -f(z)- z\overline{f}'({z})
+\frac{{\rm i}}{2\mbox{Ca}} \label{eq:eqGG}. \E 
Substitution of (\ref{eq:eqFFF}) into (\ref{eq:eqGG}) and picking $\hat{s^*} = \overline{s^*}$ so
that $g'(z)$ has no rotlet contribution (no simple poles) leads to 
\B g'(z) = -\frac{{\rm i}s^*}{(z+{\rm i})^2} + \frac{{\rm i}\overline{s^*}}{(z-{\rm i})^2} + g_{0}
 \E \vskip 0.1truein \noindent
which we rewrite as
\B g'(z) = \left [ \frac{{\rm i}s^*}{(z+{\rm i})^2} - \frac{{\rm i}\overline{s^*}}{(z-{\rm i})^2} \right ] 
-\frac{2{\rm i}s^*}{(z+{\rm i})^2} + \frac{2{\rm i}\overline{s^*}}{(z-{\rm i})^2} +
g_{0}. \label{eq:eq04}\E
As can be seen by comparison of (\ref{eq:eqFFF}) and (\ref{eq:eq04})
with (\ref{eq:eqFF}) and (\ref{eq:eqGGG}), 
the two second order poles in square brackets are associated with the two stresslets (the swimmer stresslet and its image)
and the additional second order
poles correspond to superposed potential dipole singularities (one at the swimmer position
and another at the image point). 

The constants $f_0$ and $g_0$ must be picked so that the far-field velocity condition
is satisfied.
Making use of (\ref{eq:eqFFF}) and (\ref{eq:eqGG}) in the expression for the 
velocity (\ref{eq:eqPHYS}) in the far field equation produces
\B -f_{0} + \overline{g_{0}} =  1. \label{eq:eq02b}\E \vskip 0.1truein \noindent
By balancing the constant terms in the stress condition, we can deduce
\B f_0 = -{1 \over 2} + {{\rm i} \over 4\mbox{Ca}},~~~g_0 = {1 \over 2} -{{\rm i} \over 4\mbox{Ca}}. \E

\noindent Finally, the swimmer must be stationary in the cotranslating frame. 
Hence the finite part of
\B -f(z) + z\overline{f}'(\overline{z}) + \overline{g}'(\overline{z})  \E 
at $z=-{\rm i}$ must vanish. 
This leads to
\B
0 = -\frac{\overline{s^*}}{(-2{\rm i})} - f_0 + \frac{{\rm i}s^*}{(-4)} - \frac{{\rm i}s^*}{(-4)}+ \overline{g_{0}}
\E 
and the conclusion that  
\B s^* = 2{\rm i}.\E 
Note that the strength of these singularities does not depend on the capillary number.

The solutions for $f(z)$ and $g'(z)$ just derived give the instantaneous velocity
field. We must also check that the free surface is a streamline.
It can be verified (see the discussion in the next section) that this condition is equivalent to
\B g(z) + \overline{z}f(z) = 0 \E \vskip 0.1truein \noindent
on the real axis. 
By making use of 
(\ref{eq:eqFFF}) and (\ref{eq:eqGG}) we find
\B 
\begin{split}
g(z) + zf(z) & = \frac{{\rm i} s^*}{(z+{\rm i})} - \frac{{\rm i}\overline{s^*}}{(z-{\rm i})} + zg_{0} + \frac{zs^*}{(z+{\rm i})} + \frac{z\overline{s^*}}{(z-{\rm i})} + zf_{0}\NNN
& = 2\mbox{Re}[s^*] = 0
\end{split}
\E 
which confirms that the free surface is indeed a streamline.

It still remains to satisfy the condition that ${\rm Im}[f_1]=0$ but it is easily verified
that this condition is not satisfied.  Furthermore, there are no further degrees of freedom in the
solution that will allow us to enforce it. We therefore conclude that 
steady, non-rotating solutions for a steadily translating swimmer beneath an undeformed interface
do not exist.

\section{Steady swimming beneath a deformed interface: conformal mapping}\label{sec:method}

The failure to identify a solution for a steadily translating swimmer
beneath an undeformed interface  does not preclude the existence of such solutions when the interface
deforms. We now seek such solutions.
When the interface is not flat it is no longer
obvious how to apply the method of images. 
Instead, we introduce a new solution method based on
conformal mapping. This enables us to successfully resolve the nonlinear interaction between
the point swimmer and the deformed free surface.

\begin{figure}
\begin{center}
\includegraphics[scale=0.29]{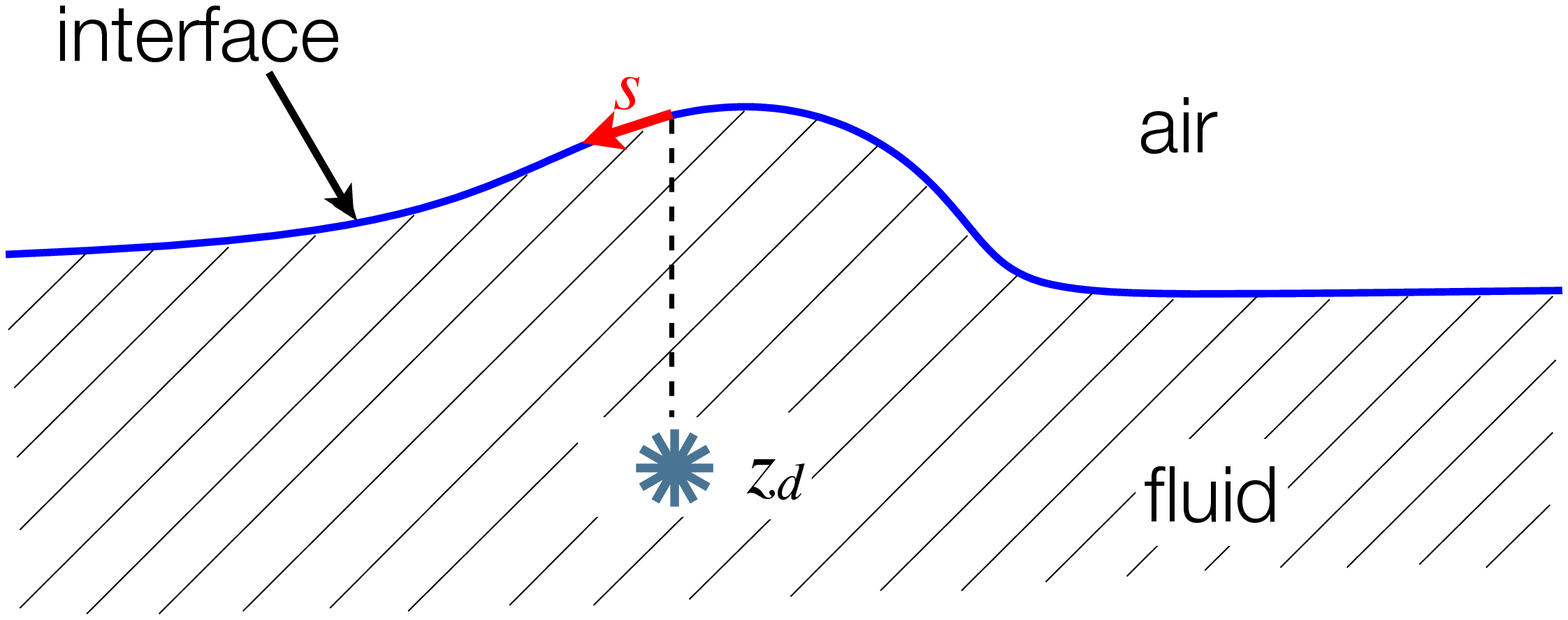}
\hskip 0.1truein\includegraphics[scale=0.29]{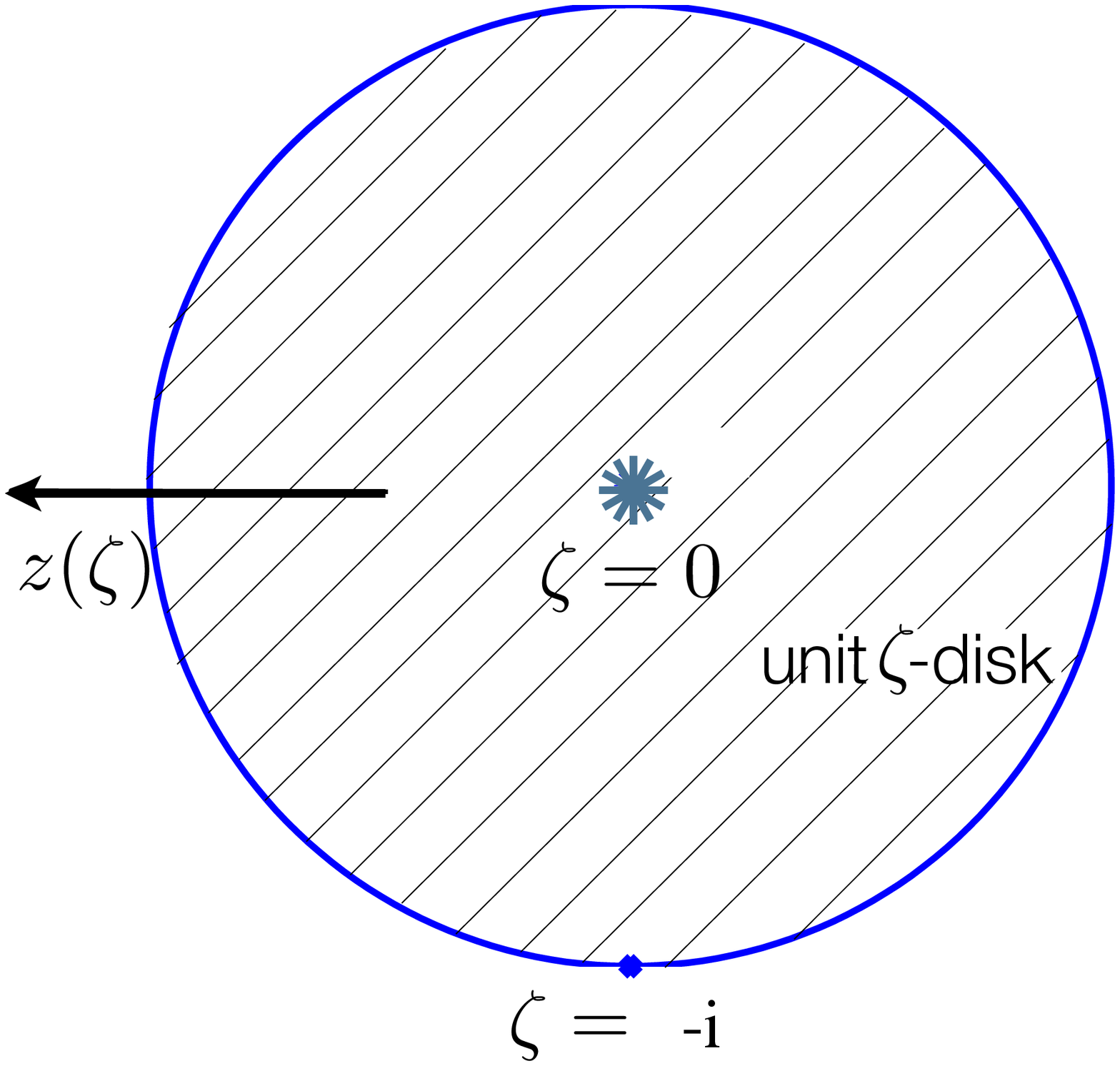}
\caption{Conformal mapping $z(\zeta)$ from the unit $\zeta$-disc to the fluid region beneath
the interface. In this mapping, $\zeta=0$ maps to the swimmer at $z=z_d$, and $\zeta=-{\rm i}$
maps to the interface at infinity. The variable $s$ \label{Fig00} denotes the arclength 
and increases as shown in the figure.}
\end{center}
\end{figure}

Consider the unit disc $|\zeta| \le 1$ in a complex parametric $\zeta$-plane.
It is known, by the Riemann mapping theorem \citep{Fokas}, that there exists a conformal
mapping $z(\zeta)$ that will map this unit disc to the fluid region beneath
the free surface. The free surface itself will be the image of the unit
circle $|\zeta|=1$ under this mapping (see Figure \ref{Fig00}).
Since the physical fluid
domain is unbounded and the free surface extends to
infinity,  there must be a simple pole of $z(\zeta)$ on the unit $\zeta$-circle.
Using a rotational degree of freedom of the Riemann mapping theorem, 
we take this pole to be at $\zeta=-{\rm i}$.
The most general form of the mapping can then be represented as
\B z(\zeta) =\frac{a}{\zeta + {\rm i}} + \sum_{k=0}^{\infty} a_{k}\zeta^k \label{eq:map}\E
where $a$ and $\{a_{k}\}$ are a (generally infinite) set of complex coefficients. 
For a one-to-one function, it must also be true that $dz/d\zeta \neq0$  
inside the unit disc. 
There are two remaining degrees of freedom in the mapping theorem which allow
us to prescribe that the singularity $z_d$ corresponding to the swimmer is
the image of $\zeta=0$, i.e,
\begin{equation}
z_d = z(0).
\end{equation}
This means that the pre-image of $z_d$ is the point $\zeta=0$.

We seek a solution
where the singularity travels uniformly with speed $U=-1$ in the $x$-direction and we again move to a co-travelling frame
in which both the swimmer and the shape of the free surface are stationary.
The kinematic condition on the interface for a steady solution in this frame is
\B {\bf u}\cdot{\bf n} = 0. \label{eq:kin}\E
If the normal vector ${\bf n} = (n_x,n_y)$, the complex normal $n_x + {\rm i} n_y$
is $-{\rm i} dz/ds$, where $s$ increases along the interface from positive infinity to negative (see Figure \ref{Fig00}).
Using this fact, (\ref{eq:kin}) can be expressed in the complex form as 
\B \mbox{Re} \left [(u+{\rm i}v){\rm i} {d\overline{z} \over ds} \right ] = 0. \label{eq:kin2} \E
This condition is equivalent to the free surface being a streamline and,
in turn, corresponds to the following condition at the free surface:
\begin{equation}
\overline{z} f(z) + g(z) = 0.
\label{eq:eq2}
\end{equation}
The easiest way to show this is to compute the derivative (with respect to arclength $s$)
of the quantity $\overline{z} f(z) + g(z)$ on the interface and then make use of
(\ref{eq:stressbalance}) and (\ref{eq:kin2}) to show that $\overline{z} f(z) + g(z)$ equals zero.

In the far-field it is required that
\begin{equation}
f(z) \to f_\infty + {f_\infty^{(1)} \over z} +\dots \qquad \mbox{and} \qquad g'(z) \to g_\infty + {g_\infty^{(1)} \over z} + \dots 
\end{equation}
\vskip 0.1truein \noindent so that, as $z \to \infty$,
\begin{equation}
u+{\rm i}v \to -f_\infty + \overline{g_\infty} + {\cal O}(|z|^{-1}).
\end{equation}
\vskip 0.1truein \noindent To satisfy the boundary conditions at infinity that the interface moves 
in the $x$-direction with unit speed we must have 
\begin{equation}
 -f_\infty + \overline{g_\infty} = 1.
 \label{eq:eq3}
 \end{equation}

\vskip 0.1truein \noindent It can be shown that
\begin{equation}
f_\infty = - \frac{1}{2} + {{\rm i} \over 4\mbox{Ca}} \qquad \mbox{and} \qquad g_{\infty} = \frac{1}{2} - 
{{\rm i} \over 4\mbox{Ca}}.
\label{eq:eq12}
\end{equation}
To see how these relations arise,
we make use of the fact that
\begin{equation}
{d z \over ds} = {{\rm i} \zeta z_\zeta \over |z_\zeta|}
\label{eq:eq6}
\end{equation}
and show that, as $\zeta \to -{\rm i}$,
\begin{equation}
- {{\rm i} \over 2} {d z \over ds}  \to {{\rm i} \over 2\mbox{Ca}} {a \over |a|}
\end{equation}
so that, using the condition (\ref{eq:stressbalance}), as $\zeta \to -{\rm i}$,
\begin{equation}
f_\infty +  \overline{g_\infty} =  {{\rm i} \over 2\mbox{Ca}} {a \over |a|}.
\label{eq:eq4f}
\end{equation}
Since, as $\zeta \to -{\rm i}$ on $|\zeta|=1$,
\begin{equation}
\begin{split}
\overline{z} = \overline{z}(\zeta^{-1}) =  {\overline{a} \zeta \over 1-{\rm i} \zeta} + \overline{a_0}
+ {\overline{a_1} \over \zeta} = {\overline{a} \over \zeta+ {\rm i}} + {\rm i} \overline{a} + \overline{a_0} 
+ {\overline{a_1} \over \zeta}
\to { \overline{a} z \over a} + ...
\end{split}
\end{equation}
it also follows, from (\ref{eq:eq2}), that
\begin{equation}
g(z) \to- { f_\infty \overline{a} z \over a},
\end{equation}
and hence
\begin{equation}
g_\infty = - { f_\infty \overline{a} \over a}.
\end{equation}
This means, from (\ref{eq:eq3}), that
\begin{equation}
-f_\infty  - { \overline{f_\infty} a \over \overline{a}}= 1.
\end{equation}
The only way for the velocity to be purely real is for $a$ to be real so that
\begin{equation}
-f_\infty  - { \overline{f_\infty}}=1.
\end{equation}
It further follows, from (\ref{eq:eq4f}), that
\begin{equation}
f_\infty - \overline{f_\infty} =  {{\rm i} \over 2\mbox{Ca}},
\label{eq:eq4}
\end{equation}
leading to (\ref{eq:eq12}).

\vskip 0.1truein \noindent Together with the stress condition (\ref{eq:stressbalance}), the kinematic condition (\ref{eq:kin}) can 
be written as
\begin{equation}
{\rm Re}\biggl [ 2 f {\rm i} \overline{\left ({dz \over ds}\right )} \biggr ] = {1 \over 2\mbox{Ca}}.
\end{equation}
We now introduce the composed functions
\begin{equation}
F(\zeta) \equiv f(z(\zeta)),~~~ G(\zeta) \equiv g(z(\zeta)),
\end{equation}
which can be used in the kinematic condition together with (\ref{eq:eq6}) to give
\begin{equation}
{\rm Re}\biggl [ {2 F(\zeta) \over \zeta z_\zeta} \biggr ] = {1 \over 2\mbox{Ca} |z_\zeta|},
\end{equation}
where $z_\zeta(\zeta) \equiv {dz/d\zeta}$.
Since $f(z)$ is required to have a simple pole at $z_d$,
$F(\zeta)$ must have a simple pole at $\zeta=0$. Therefore  
\begin{equation}
F(\zeta) = {F_d \over \zeta} + F_0 + {\cal O}(\zeta)
\end{equation}
for some constant $F_{d}$ (to be determined). 
Therefore, consider
\begin{equation}
\begin{split}
{\rm Re}\biggl [ {F(\zeta) \over \zeta z_\zeta} - {D \over \zeta^2} - {C \over \zeta}
\biggr ]  = {1 \over 4\mbox{Ca} |z_\zeta|}-{\rm Re} \biggl [ {D \over \zeta^2} + {C \over \zeta}
\biggr ]  
= 
{1 \over 4 \mbox{Ca}|z_\zeta|}-{\rm Re}\biggl [  {\overline{D} \zeta^2} + {\overline{C} \zeta}
\biggr ] .
\end{split}\label{eq:eq5f}
\end{equation}
For suitable choices of constants $C$ and $D$
the function in the square brackets on the left hand side of 
(\ref{eq:eq5f}) is analytic everywhere inside the unit $\zeta$-disc. 
Equation (\ref{eq:eq5f}) therefore gives the real part of an analytic function 
on the boundary of the disc. The Poisson integral formula can be used to give
\begin{equation}
F(\zeta) = \zeta z_\zeta(\zeta) \biggl [ I(\zeta,\mbox{Ca}) + {D \over \zeta^2} + {C \over \zeta}
- {\overline{D} \zeta^2} - {\overline{C} \zeta} +{\rm i} b \biggr ], \label{eq:eq6f}
\end{equation}
where $b$ is some real constant and
\begin{equation}
I(\zeta,\mbox{Ca}) = {1 \over 8 \pi {\rm i}\mbox{Ca}} \oint_{|\zeta'|=1} {d \zeta' \over \zeta'}
{\zeta'+ \zeta \over \zeta'-\zeta} {1 \over |z_\zeta(\zeta')|}\cdot
\label{eq:eq7}
\end{equation}
Since
$z_\zeta(\zeta)$ has a second-order pole at $\zeta=-{\rm i}$, it is necessary that the quantity inside the brackets in expression (\ref{eq:eq6f}), and its derivative, vanish at $\zeta = -{\rm i}$ so that near $\zeta=-{\rm i}$
\begin{equation}
I(\zeta,\mbox{Ca}) + {D \over \zeta^2} + {C \over \zeta}
- {\overline{D} \zeta^2} - {\overline{C} \zeta} + {\rm i} b = A (\zeta+{\rm i} )^2 + \dots \label{eq:6ff}
\end{equation}
for some complex constant $A$ which is related to $f_\infty$ by
\begin{equation}
{\rm i} a A = f_\infty.
\label{eq:eq12aa}
\end{equation}
This means that
\begin{equation}
I(-{\rm i} ,\mbox{Ca}) - D+ \overline{D} +{\rm i}  C + {\rm i} \overline{C} + {\rm i} b = 0
\label{eq:eq10},
\end{equation}
and
\begin{equation}
I_\zeta(-{\rm i},\mbox{Ca}) + {2 {\rm i} D } +{C}
+ 2 \overline{D} {\rm i} - \overline{C} = 0
\label{eq:eq11},
\end{equation}
while 
\begin{equation}
A = {1 \over 2} \biggl [ I_{\zeta \zeta}(-{\rm i} ,\mbox{Ca}) + 6 D - 2 C{\rm i}  - 2 \overline{D} \biggr ].
\label{eq:eq15}
\end{equation}
Since $I(-{\rm i},\mbox{Ca})$ is purely imaginary,
(\ref{eq:eq10})
is a single real condition, namely, an equation for the real parameter
$b$ in terms of $C$ and $D$.

If the map $z(\zeta)$ and 
the constants $b, C$ and $D$ are known,
(\ref{eq:eq7}) gives an explicit expression for $F(\zeta)$.
Condition
(\ref{eq:eq2}) then provides the following expression
for $G(\zeta)$:
\begin{equation}
G(\zeta) = - \overline{z}(\zeta^{-1}) F(\zeta). \label{eq:eq7f}
\end{equation}
This condition is very revealing: it shows that the singularities of $G(\zeta)$
dictate the singularities of the conformal mapping function $z(\zeta)$.
In particular, it shows that if $G(\zeta)$ has a pole singularity at $\zeta=0$ then
the functional form of the mapping function is finite.
To see this suppose, for example, that $z(\zeta)$ has the special truncated form
\B z(\zeta)  = \frac{a}{\zeta + {\rm i}} + a_{0} + \sum_{k=1}^{n} a_{k} \zeta^k  \label{eq:eqmap22}, \E
where $n\ge 1$ is some positive integer. 
Then
\B \overline{z}(\zeta^{-1})  
= {\overline{a} \zeta \over 1 - {\rm i} \zeta} + \overline{a_{0}} + 
\sum_{k=1}^{n} {\overline{a_{k}} \over \zeta^k}\cdot \label{eq:eqmap22a}\E
Since
$F(\zeta) = {\cal O}(\zeta^{-1})$ as $\zeta \to 0$,
it follows from equation (\ref{eq:eq7f}) and (\ref{eq:eqmap22a})
that, near $\zeta = 0$, $G(\zeta)$ has the form 
\B G(\zeta) = \frac{G_{-(n+1)}}{\zeta^{n+1}} + \frac{G_{-n}}{\zeta^n} + \dots  + \frac{G_{-1}}{\zeta } + G_{0}  + G_{1} \zeta + \dots \E
implying that, near $z_{d}$,
\B g(z) = \frac{g_{-(n+1)}}{(z-z_{d})^{n+1}} + \mbox{higher order terms}. \E
Since our swimmer model means that $g(z)$ must have the form given in (\ref{eq:eqGForm})
it is clear that we must pick
$n=1$.
The conformal map then has the form 
\B
z(\zeta) = {a \over \zeta+{\rm i}} + a_0 + a_1 \zeta,
\label{eq:eqMAP2}
\E
where $a, a_0$ and $a_1$ are
complex numbers.

\begin{figure}
\begin{center}
\includegraphics[scale=0.4]{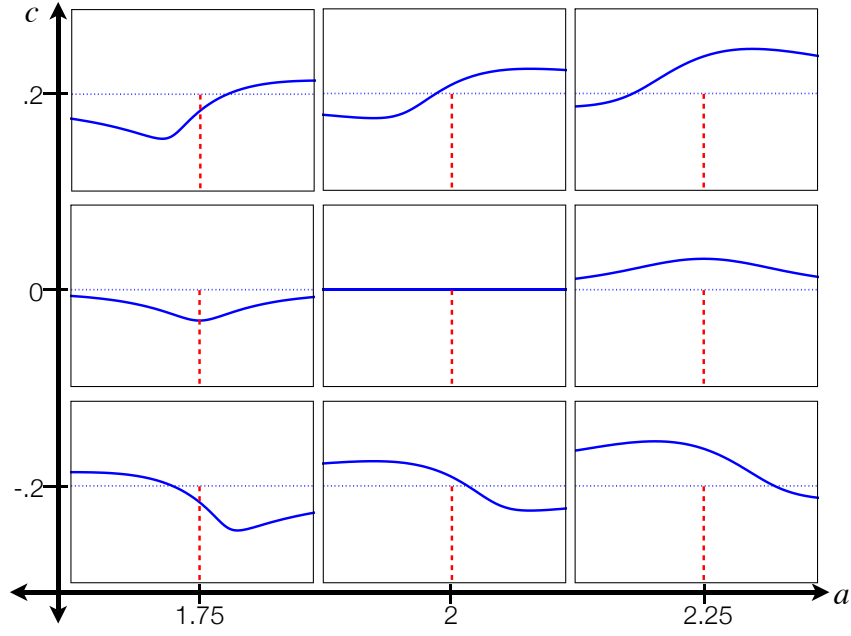}
\caption{The geometrical significance of parameters $a$ and $c$. The flat interface
 corresponds to $a=2, c=0$.  Increasing (decreasing) $a$ above $2$ with $c=0$ (along the horizontal axis) moves the interface
 up (down) in a left-right symmetric fashion.  Non-zero values of $c$ (along the vertical axis) introduce
 left-right asymmetry. \label{FigILL}}
 \end{center}
\end{figure}

In addition, there are geometrical constraints on the conformal mapping parameters.
Suppose that the singularity is at $z_d=-{\rm i}$. This implies that $z(0)=-{\rm i}$ 
and therefore that
\B a_0 = (a-1){\rm i}. \E
Furthermore, the condition that the interface tends to $y=0$ implies
\B {\rm Re}[a_1] = {a \over 2} -1. \E
Hence, the mapping function can be expressed as
\begin{equation}
z(\zeta) = {a \over \zeta+{\rm i}} +{\rm i}(a-1) + \left ( {a \over 2} -1 + {\rm i}c \right)\zeta,
\label{eq:eqmap}
\end{equation}
where $a$ and $c$ are two real parameters.
These two parameters have the following geometrical interpretations.  When $a=2$ with $c=0$ the
interface is flat; increasing $a$ above 2 corresponds to a symmetric {\em upward}
deformation of the interface; decreasing $a$ below 2 leads to a symmetric {\em downward}
deformation. Changing the value of $c$ away from zero introduces
left-right {\em asymmetry} of the free surface deformation about a vertical axis through the
swimmer, as illustrated in Figure \ref{FigILL}.

The velocity of the swimmer in the co-moving frame is given by the finite part of the
expression $u+{\rm i}v$ at $z= -{\rm i}$ and, for a steady solution,
this must vanish.
If
\B f(z) = \frac{s^*}{z+{\rm i}} + f_0 + f_{1}(z+{\rm i}) + \dots, ~~~
g(z) =  {q^* \over(z+{\rm i})^2} + {(- {\rm i} s^* + d^*) \over (z+{\rm i})}  +
 g_0 + g_{1}(z+{\rm i}) + \dots\E
the finite part of $u+{\rm i}v$ at $z=-{\rm i}$ is given by
\B -f_0 -{\rm i} \overline{f_{1}} + \overline{g_{1}} = 0.\label{eq:eq8f}\E
The terms $f_0$, $f_1$ and $g_1$ are given (using the residue theorem) as 
\B
\begin{split}
f_{0} = \frac{1}{2\pi {\rm i}} \oint_{\Gamma}&\frac{F(\zeta)}{(z(\zeta)+{\rm i})} \frac{dz}{d\zeta}~d\zeta, ~~~~
f_{1}  = \frac{1}{2\pi {\rm i}} \oint_{\Gamma}\frac{F(\zeta)}{(z(\zeta) +{\rm i})^2} \frac{dz}{d\zeta}~d\zeta, \\
g_{1} &= - \frac{1}{2 \pi{\rm i}} \oint_{\Gamma}\frac{\overline{z}(1/\zeta)F(\zeta)}{(z(\zeta)+{\rm i})^2} \frac{dz}{d\zeta}~d\zeta
\end{split} \label{eq:eq9f}
 \E
where $\Gamma$ is any simple closed curve surrounding $\zeta=0$. 
Note that with $F(\zeta)$ determined from equation (\ref{eq:eq6f}), $G(\zeta)$ from (\ref{eq:eq7f}) and $f_0$, $f_{1}$ and 
$g_1$ from (\ref{eq:eq9f}), condition (\ref{eq:eq8f}) 
is a {linear} equation for constants $C$ and $D$. 
It should also be pointed out that $f_0, f_1$ and $g_1$ can, in principle, be found in analytical
form but the algebra involved is lengthy and it is easier to make use of the integral expressions
(\ref{eq:eq9f}) to compute these quantities.

If the values of $a,c$ and $U$ are specified,
$C$ and $D$ can  be found by simultaneously solving
\begin{equation}
-f_0 - {\rm i} \overline{f_1} + \overline{g_1} = 0,
\label{eq:eq16c}
\end{equation}
which is a complex equation,
together with the real equation (\ref{eq:eq11}), i.e.,
\begin{equation}
I_\zeta(-{\rm i},{\rm Ca}) + {2 {\rm i} D } +{C}
+ 2 \overline{D} {\rm i} - \overline{C} = 0,
\label{eq:eq16b}
\end{equation}
and the real equation
\begin{equation}
-{1\over 2} = {\rm Re} \biggl [
{{\rm i} a \over 2} \biggl [ I_{\zeta \zeta}(-{\rm i},{\rm Ca}) + 6 D - 2 C{\rm i} - 2 \overline{D} \biggr ]
\biggr ].
\label{eq:eq16a}
\end{equation}
With constants $C$ and $D$ determined in this way, functions $F(\zeta)$ and
$G(\zeta)$ (and hence the flow field) are fully determined in analytical
form.
The stresslet, dipole and quadrupole strengths can then be readily computed to be
\begin{equation}
\begin{split}
s^* &= D \left ({3 a -2 \over 2}  + {\rm i} c \right )^2, \\
q^* &= - D \left ({a-2 \over 2} - {\rm i} c \right )  \left ({3 a -2 \over 2} + {\rm i} c \right )^2, \\
d^* &=  \left ({3 a -2 \over 2}  + {\rm i} c \right ) \left [ {\rm i} D a \left (
- {a \over 2} +3 + 5 {\rm i} c  \right )
- C \left ({a -2 \over 2} - {\rm i} c \right )  \left ({3 a -2 \over 2}  + {\rm i} c \right ) \right ].
\end{split}
\end{equation}
As a check on the solution scheme, it was verified that when $a=2, c=0$ (so that the
interface is flat) we retrieved the values
\begin{equation}
s^* = 2 {\rm i}, ~~q^*=0,~~d^*=-4,
\end{equation}
which are the values obtained in \S \ref{MI} using the method
of images.

We have thus found that there is a two-parameter family of
solutions, parametrized by $a$ and $c$, 
for a force-free (but not necessarily non-rotating) point swimmer  -- characterized by a singularity description of the form (\ref{eq:eqGForm}) --
steadily translating beneath a deformed free surface with speed $U=-1$.

It remains, however, to enforce the condition ${\rm Im}[f_1]=0$.
In contrast to the case when the interface is assumed to be flat, we now have
additional freedoms (in the choice of $a$ and $c$) to enforce this condition.
Physically, this means that possible equilibrium solutions exist within a two-parameter
family of possible interface shapes.
With a two-parameter family of possible solutions, and 
only a single
additional requirement, we choose to specify $a$ and see if there are any corresponding
values of $c$ such that ${\rm Im}[f_1]=0$. It must also then be checked {\em a posteriori} that
the map (\ref{eq:eqmap}) with these values of $a$ and $c$
are one-to-one (univalent) mappings
to the fluid region.
It has been found that such solutions do indeed exist. They are described
in the next section.

\section{Characterization of the steady solutions \label{eff}}

For a given value of $a$, admissible values of $c$ are found by insisting that
\begin{equation}
{\rm Im}[f_1]=0.
\label{eq:eqVORT}
\end{equation}
The nature of our formulation is such that 
the only stage in which we are required to solve a nonlinear equation is in satisfying (\ref{eq:eqVORT}).
Newton's method is used to satisfy this condition by fixing a value of $a$,
taking a guess for $c$ and then computing the associated functions $F(\zeta)$ and $G(\zeta)$
together with the values of $f_0, f_1$ and $g_1$. The value of $c$ is then updated iteratively
until condition (\ref{eq:eqVORT}) is satisfied.
It is then checked {\em a posteriori} that the resulting conformal mapping is a one-to-one
mapping from the interior of the unit $\zeta$-disc to the fluid domain.

\begin{figure}
\begin{center}
\includegraphics[scale=0.35]{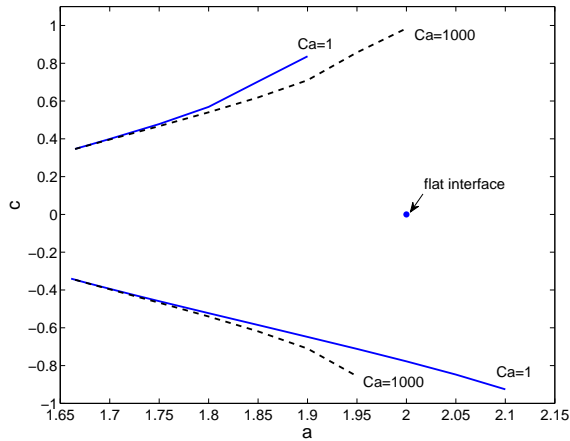}
\caption{The two solution branches: the values of $c$ against $a$ are plotted
for ${\rm Ca}=1$ and $1000$. 
The upper branches have positive values of $c$, the lower branches have
negative values of $c$. The flat interface parameters are also indicated
to show that the actual solution branches are disconnected from it.
\label{NewF1}}
\end{center}
\end{figure}

\begin{figure}
\begin{center}
\includegraphics[scale=0.4]{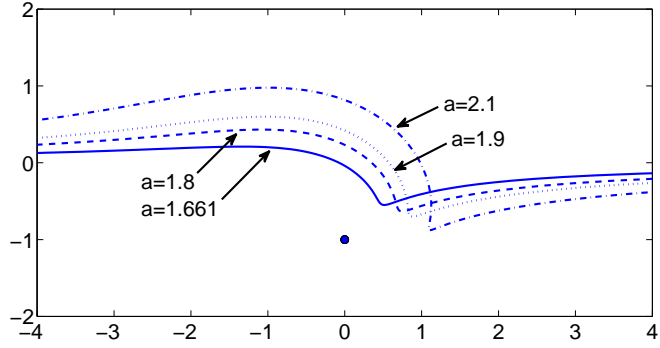}
\end{center}
\caption{Free surface profiles (with ${\rm Ca}=1$) for four different $a$ in the range of existence of solutions for the lower solution branch in Figure \ref{NewF1} corresponding to $c$ negative. \label{NewF2}}
\end{figure}

\begin{figure}
\begin{center}
\includegraphics[scale=0.35]{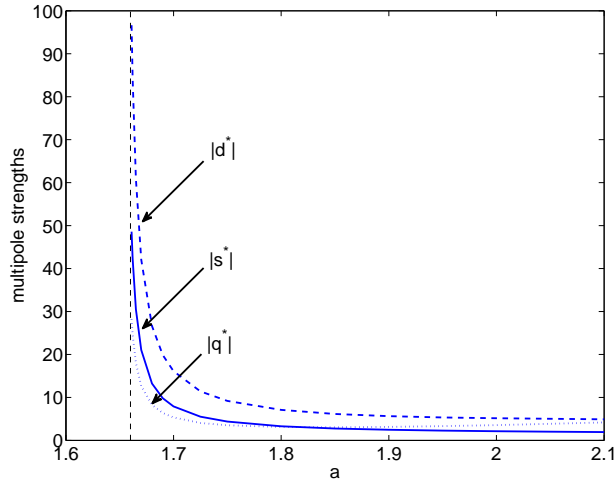}
\caption{Multipole strengths for ${\rm Ca}=1$ in the range of
existence of solutions for the lower solution branch with $c$ negative.
\label{NewF3}}
\end{center}
\end{figure}

Figures \ref{NewF1} shows a graph of $c$ against $a$ when
${\rm Ca}=1$ and $Ca=1000$. Focussing on the case ${\rm Ca}=1$ first
it is clear that there are two solution branches: 
for certain values of $a$ (but not all) two possible values of 
$c$ giving physical admissible
solutions are found.
This is not surprising given that $a$ and $c$ are simply mathematically convenient
parameters that are related to physically meaningful parameters (e.g. the multipole strengths)
via highly nonlinear
relations.

The lower solution branch corresponds to 
$c < 0$: Figure \ref{NewF2} shows typical profiles throughout the range of existence
of this branch
and reveals that the point of highest
curvature is to the right of the swimmer.

Figure \ref{NewF3} shows the magnitudes of the multipole strengths
as a function of $a$ for this branch of solutions. Interestingly, as $a \to 1.66$ which
is the lowest $a$-value for which solutions are found to exist on this branch the free surface profile
tends to a configuration in which the point of highest curvature tends to a finite value but
the multipole strengths appear to grow without bound (hence the vertical asymptote shown
in Figure \ref{NewF3}). On the other hand, as $a$ tends to the upper value for
which solutions exist the value of the maximum
curvature of the interface appears to grow
without bound while the values of the multipole strengths appear to asymptote to
finite values.

Figures \ref{NewF4} and \ref{NewF5} show similar graphs for the upper branch
for which $c$ is positive. Now the point of highest curvature on the interface is to the left
of the swimmer position. The qualitative behaviour of the solution branch is very similar,
however, to the lower branch just investigated. At one end of the range of existence of solutions
the maximum curvature of the interface tends to a finite value with the multipole strengths
growing without bound; at the other end of the range, the maximum curvature of the interface
grows without bound with the multipole strengths levelling off to well-defined values.

\begin{figure}
\begin{center}
\includegraphics[scale=0.4]{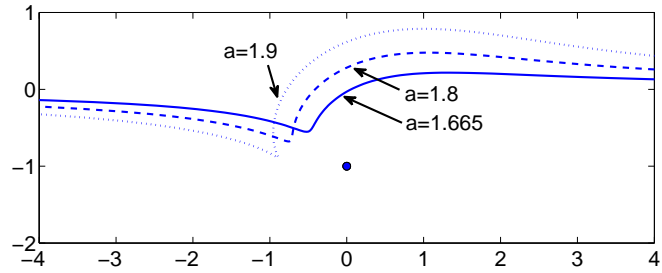}
\caption{Free surface profiles (with ${\rm Ca}=1$) for three different $a$ in the range of
existence of solutions for the upper solution branch in Figure \ref{NewF1} corresponding to $c$ positive.
\label{NewF4}}
\end{center}
\end{figure}

\begin{figure}
\begin{center}
\includegraphics[scale=0.35]{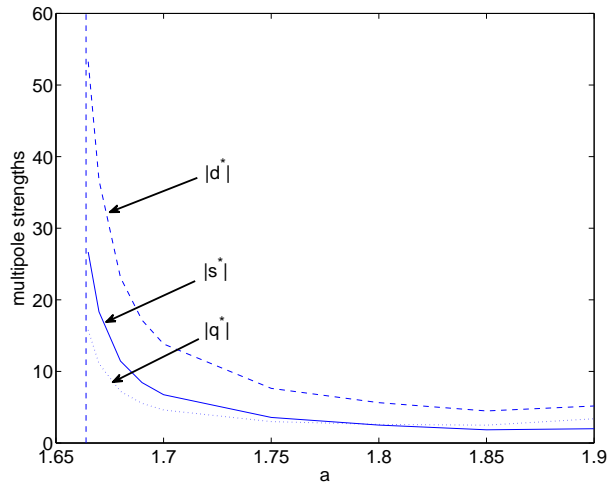}
\caption{Multipole strengths for ${\rm Ca}=1$ in the range of
existence of solutions for the upper solution branch with $c$ positive.
\label{NewF5}}
\end{center}
\end{figure}

\begin{figure}
\begin{center}
\includegraphics[scale=0.32]{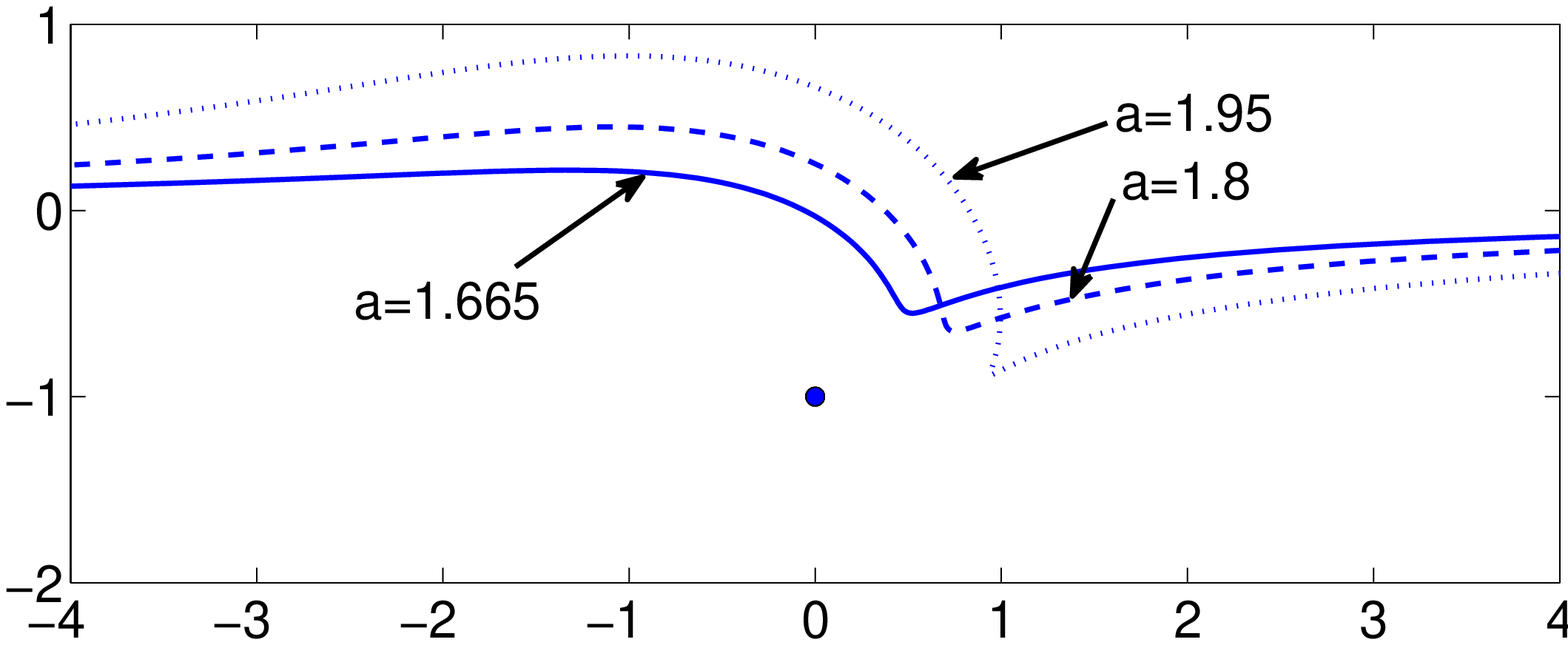}
\includegraphics[scale=0.28]{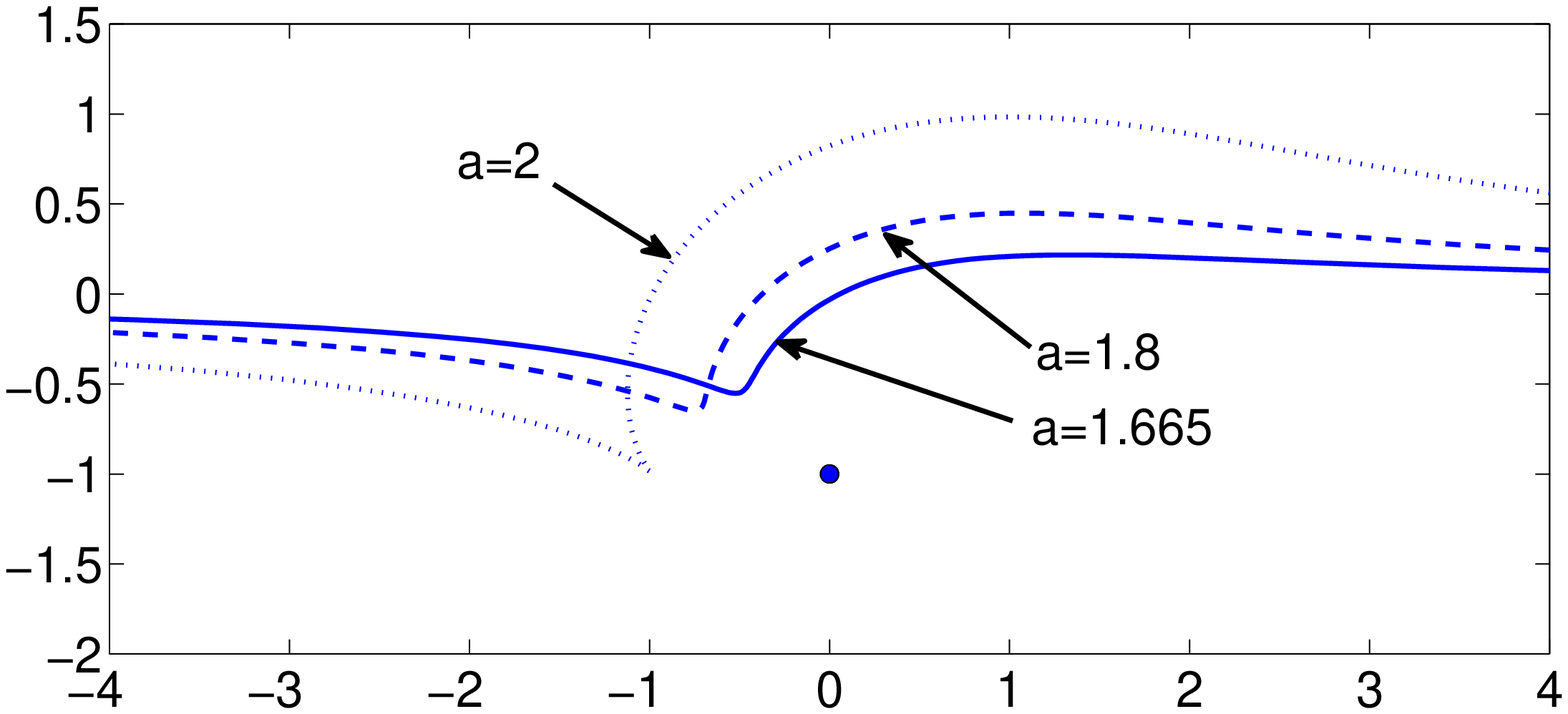}
\caption{Free surface profiles for the two solution branches with ${\rm Ca}=1000$ (left: lower branch; right: upper branch).
Shapes are qualitatively similar to the ${\rm Ca}=1$ case 
\label{NewF7} (Figs.~\ref{NewF2} and \ref{NewF4}).}
\end{center}
\end{figure}

The above results appear to be only weakly dependent on the capillary number ${\rm Ca}$.
Figure \ref{NewF1} also shows $c$ against $a$ in the case  ${\rm Ca}=1000$ and once again
two distinct solutions branches are found (disconnected from the flat state solution).
One solution branch has $c < 0$, a second has $c >0$.
The free surface profiles for the two solution branches are depicted in Figure \ref{NewF7}
and are qualitatively similar to the ${\rm Ca} =1$ case.

It appears therefore that modes of steady translation are possible and that they are
associated with a range of free surface profiles. What {is} clear from the solutions
is that the steady motion requires significant deformations of the surface.
In other words, the nonlinear interaction between the swimmer and the deformability of the
interface
is crucial for steady swimming.

It is natural to ask which of the family of steady solutions 
might be realized in practice or be energetically preferable.
Usually such matters are decided based on some definition of swimmer efficiency.
An example is to minimize the energy dissipation associated with a displacement
of the swimmer by unit distance. 
Unfortunately, owing to the fact that the model used here involves introducing singularities
in the flow then the associated dissipation rates are unbounded (however, some progress
in this direction has recently been made by \cite{Lee2009} where use is made of the
exact solution class found here).
Nevertheless, it is intuitively sensible that the particular biological
characteristics of any given swimmer will
mean that there is an upper bound on the strength of the far-field multipoles
in any effective singularity description of the swimmer. 
Moreover, it is also sensible to suppose that the rate of energy dissipation will be
smaller when the multipole strengths are smaller.
Figures \ref{NewF3}
and \ref{NewF5} show that the multipole strengths are least when the values of $a$ are greatest. 
As shown in Figures \ref{NewF2}
and \ref{NewF4} 
it is for these higher values of
$a$ that the free surface tends towards a near-cusped interfacial shape (either in front of, or behind, the
swimmer).
These considerations suggest
that efficient swimmers in steady motion will be inclined to produce
free surface deformations exhibiting a region of high curvature.

\section{Conclusion}\label{sec:concl}

In this paper we have used complex analysis to identify a new mode 
of low-Reynolds swimming, 
namely organisms or devices that generate flow disturbances which 
exploit the deformation of a nearby free surface to swim steadily parallel to the interface. 
The resulting surface shapes are illustrated in Figures \ref{NewF2}, \ref{NewF4} and \ref{NewF7}. The two solution branches we obtain, displayed in Figure \ref{NewF1} are seen to be disconnected from the flat state solution. The interaction between the swimmer's disturbance to the flow, the free surface, 
and the subsequent swimmer's dynamics is highly nonlinear, and it is likely to be  difficult to derive the solutions presented here
by a weakly nonlinear perturbation analysis about the flat state.
It is therefore of mathematical interest that
by employing complex variable methods and a singularity model 
we have found these nonlinear interactions to be mathematically tractable.

The swimmer model we have used is motivated by a simple treadmilling circular
motion of the kind first considered by \citep{Blake1972}.
In principle the mathematical techniques used here are extendible to more detailed and realistic singularity representations. Furthermore it should be possible to extend this study by using the method of matched asymptotic
expansions  to match an ``inner'' flow generated by
a small finite-area swimmer with an ``outer''
solution in which the flow generated by that swimmer interacts with the free surface.
The assumptions of such an analysis would be that the swimmer size is small compared
to its distance from the interface so that, at leading order in an expansion based on 
a parameter expressing this ratio of scales, the swimmer would look like a point singularity.
The solutions found here would then
serve as the outer solutions in such a scheme.
\cite{Antanovskii} has used precisely such a strategy of matched asymptotics in his two-dimensional
complex variable model of a deformable bubble in the flow field generated by Taylor's
four-roller mill.

The bearing of our two-dimensional analysis on the physics of actual organisms employing a free surface 
for propulsion is not clear. Other models for free surface locomotion have been proposed \citep{Lee2008} also involving free surface interaction, but they were restricted to small-amplitude motion. Here, using a fully nonlinear analysis,  we have unveiled an essential mechanism for the steady locomotion of a swimmer near a free capillary surface. In particular, our analysis has demonstrated the  essential role played by interface deformability, and hinted at the significance  of regions with high surface curvature.

It is likely that, as in the case of swimming near a no-slip wall, 
the generic locomotion mechanism of an organism near a free surface will have a more complicated spatio-temporal structure. Indeed, free surface deflection associated with unsteady
undulatory waves propagating along the foot of a water snail have been observed in
practice.
In future work it may be interesting to study the full unsteady dynamics of a point swimmer
near a free capillary surface. The geometrical complexities even in that simplified case
would no doubt require a fully numerical investigation. 
An analysis of the stability of the steadily translating
equilibria just identified is also of interest but is left for a future study.

 \vskip 0.3truein

This work was funded in part by the US National Science Foundation 
through grants CTS-0624830 (EL and AEH) and CBET-0746285 (EL).
DC acknowledges the support of an EPSRC Advanced Research Fellowship
as well as the hospitality of the Department of Mathematics at MIT
where this work was initiated. OS also acknowledges support from an EPSRC studentship.

\bibliographystyle{unsrt}
\bibliography{PtS}

\end{document}